\documentclass[preprint,superscriptaddress,onecolumn,amsmath,amssymb]{revtex4-1}

\usepackage{graphicx}
\usepackage{float}

\begin{document}
\title{Ubiquitous antinodal quasiparticles and deviation from simple \textit{d}-wave form in underdoped Bi-2212}
\author{I. M. Vishik}
\affiliation {Massachusetts Institute of Technology, Department of Physics, Cambridge, MA, 02139, USA}
\author{Makoto Hashimoto}
\affiliation {Stanford Synchrotron Radiation Lightsource, SLAC National Accelerator Laboratory, Menlo Park, California 94025, USA}
\author{W. S. Lee}
\affiliation {Stanford Institute for Materials and Energy Sciences, SLAC National Accelerator Laboratory, 2575 Sand Hill Road, Menlo Park, CA 94025, USA}
\author{T. P. Devereaux}
\affiliation {Stanford Institute for Materials and Energy Sciences, SLAC National Accelerator Laboratory, 2575 Sand Hill Road, Menlo Park, CA 94025, USA}
\author{Z. X. Shen}
\affiliation {Stanford Institute for Materials and Energy Sciences, SLAC National Accelerator Laboratory, 2575 Sand Hill Road, Menlo Park, CA 94025, USA}
\affiliation{Geballe Laboratory for Advanced Materials, Stanford University, Stanford, CA 94305, USA}

\date{\today}

\begin{abstract}
The momentum dependence of the superconducting gap in the cuprates has been debated, with most experiments reporting a deviation from a simple $d_{x^2-y^2}$ form in the underdoped regime and a few experiments claiming that a simple $d_{x^2-y^2}$ form persists down to the lowest dopings.  We affirm that the superconducting gap function in sufficiently underdoped Bi$_2$Sr$_2$CaCu$_2$O$_{8+\delta}$ (Bi-2212) deviates from a simple \textit{d}-wave form near the antinode.  This is observed in samples where doping is controlled only by oxygen annealing, in contrast to claims that this effect is only seen in cation-substituted samples.  Moreover, a quasiparticle peak is present at the antinode down to p$=$0.08, refuting claims that a deviation from a simple \textit{d}-wave form is a data analysis artifact stemming from difficulty in assessing a gap in the absence of a quasiparticle.  
\end{abstract}

\maketitle
\textbf{Significance Statement:}\textit{The origin of high temperature superconductivity in the cuprates is an enduring question in condensed matter physics.  A key difficulty is that the electronic phase existing at temperatures above the superconducting transition temperature is poorly understood.  This ‘pseudogap’ phase is alternately attributed to fluctuating superconductivity or to a state which is distinct from superconductivity, and these competing scenarios have different implications for the origin of superconductivity.  We report evidence in support for the latter scenario, which manifests in the angular dependence of the superconducting gap.}

\section{Introduction}
Angle-resolved photoemission spectroscopy (ARPES) can measure the spectral gap in the superconducting state to great precision, but yet there are still disputes about the details of the momentum dependence.  In overdoped Bi-2212, there is wide agreement that the superconducting gap function follows a simple \textit{d}-wave form, expressed at 0.5$*|\cos(k_x)-\cos(k_y)|$ or $\cos(2\theta)$ \cite{Vishik:PNAS, anzai:RelationNodalAntinodalGap, Zhao:UniversalFeatures}.  We use the former convention in this paper.  In the underdoped regime, there are experiments that find that the gap function deviates from a simple \textit{d}-wave form \cite{Tanaka:twoGapARPES_dopingDep,Lee:twoGapARPES_TDep,Vishik:PNAS, anzai:RelationNodalAntinodalGap}, with the antinodal region exhibiting a larger gap than implied by the simple \textit{d}-wave trend in the near nodal region.  This is reported in Bi-2212, (Bi,Pb)$_2$(Sr,La)$_2$CuO$_{6+\delta}$ (Bi-2201) \cite{Kaminski:2gapBi2201}, La$_{2-x}$Sr$_x$CuO$_{4+\delta}$ (LSCO) \cite{SCGapLSCO:Teppei,He:LBCO_one_eigth}, the inner plane of Bi-2223 which is more underdoped than the outer planes \cite{Ideta:EnhancedSC_Gaps}, and YBa$_2$Cu$_3$O$_y$ (YBCO) \cite{Nakayama:DopingDepYBCO}.  Recently, Zhao \textit{et al} have reported that the reported deviation from a simple \textit{d}-wave form in Bi-2212 is an artifact of cation substitution which is used to achieve stable underdoping \cite{Zhao:UniversalFeatures}.  They claim that cation substitution (Dy or Y) on the Ca site suppresses the antinodal quasiparticle, and when antinodal quasiparticles are present, the gap function always follows a simple \textit{d}-wave form.  They also claim that without cation substitution, the gap function always follows a simple \textit{d}-wave form.

In the ARPES literature, a deviation of the superconducting gap function from a simple \textit{d}-wave form is often taken as evidence that the pseudogap coexists with superconductivity below T$_c$, which implies that it is a distinct phase.  However, the purpose of this paper is not to discuss the relationship between the pseudogap and superconductivity.  Rather, we simply aim to clarify the experimental facts.  We show data demonstrating a deviation from a simple \textit{d}-wave form in a sample which was underdoped only by oxygen annealing.  Then, we demonstrate that antinodal quasiparticles are clearly observed in samples with various compositions down to a T$_c$ of 50K (p$\approx$0.08), and that the gap function deviates from a simple \textit{d}-wave form in underdoped samples with a T$_c$ of 75K or smaller.  We explore various reasons for the discrepancy between our result and that of Zhao \textit{et al} including matrix elements and resolution.

\begin{table*}[!]

\begin{tabular}{ l c c c c c r }

Sample & Composition & Photon energy & Cut geometry & Resolution & Temperature & Figure \\
  \hline
  UD50 & Bi$_2$Sr$_2$(Ca,Y)Cu$_2$O$_{8+\delta}$ & 19eV & $\Gamma$Y (2nd BZ) & 14 meV & 10K & \ref{Fig 2: Ubiquitous QPs}\\
  UD55 & Bi$_2$Sr$_2$(Ca,Dy)Cu$_2$O$_{8+\delta}$ & 18.4, 19, 21, 22.7 eV & $\Gamma$M & 8 meV &9K & \ref{Fig 3: matrix elements} \\
  UD65 & Bi$_{2+x}$Sr$_{2-x}$CaCu$_2$O$_{8+\delta}$ & 22.7 & $\Gamma$M & 8 meV & 12K & \ref{Fig 2: Ubiquitous QPs} \\
  UD75 & Bi$_2$Sr$_2$CaCu$_2$O$_{8+\delta}$ & 22.7eV & $\Gamma$M & 7 meV & 10K & \ref{Fig 1: UD75} \\
  UD92 & Bi$_2$Sr$_2$CaCu$_2$O$_{8+\delta}$ & 22.7eV & $\Gamma$M & 7 meV & 10K & \ref{Fig 2: Ubiquitous QPs}\\
  \hline
\end{tabular}
\label{Table 1}
\caption{Summary of samples discussed in this work with their composition and experimental setup. $\Gamma$Y refers to cuts taken parallel to the (0,0)-($\pi$,$\pi$) line and $\Gamma$M refers to cuts taken parallel to ($\pi$,0)-($\pi$,$\pi$). UD(OD) refers to underdoped (overdoped) samples and the number which follows gives the T$_c$. }
\end{table*}

ARPES experiments were performed at Stanford Synchrotron Radiation Lightsource (SSRL).  Samples were cleaved \textit{in situ} at the measurement temperature (T$<$T$_c$) at a pressure better than 5$\times$10$^{-11}$ Torr.  The Fermi energy, E$_F$, was determined from the Fermi edge of polycrystalline gold which is electrically connected to the sample.  Table \ref{Table 1} shows the composition and experimental configuration for all of the samples presented in this paper.  Most samples were measured with a SCIENTA R4000 analyzer (angular resolution $\approx0.1^\circ$), except for UD50 which was measured with a SCIENTA SES-200 analyzer (angular resolution $\approx0.3^\circ$).  Energy resolutions are also listed in Table \ref{Table 1}.  
\begin{figure*}[!]
\includegraphics [type=eps,ext=.eps,read=.eps,clip, width=6 in]{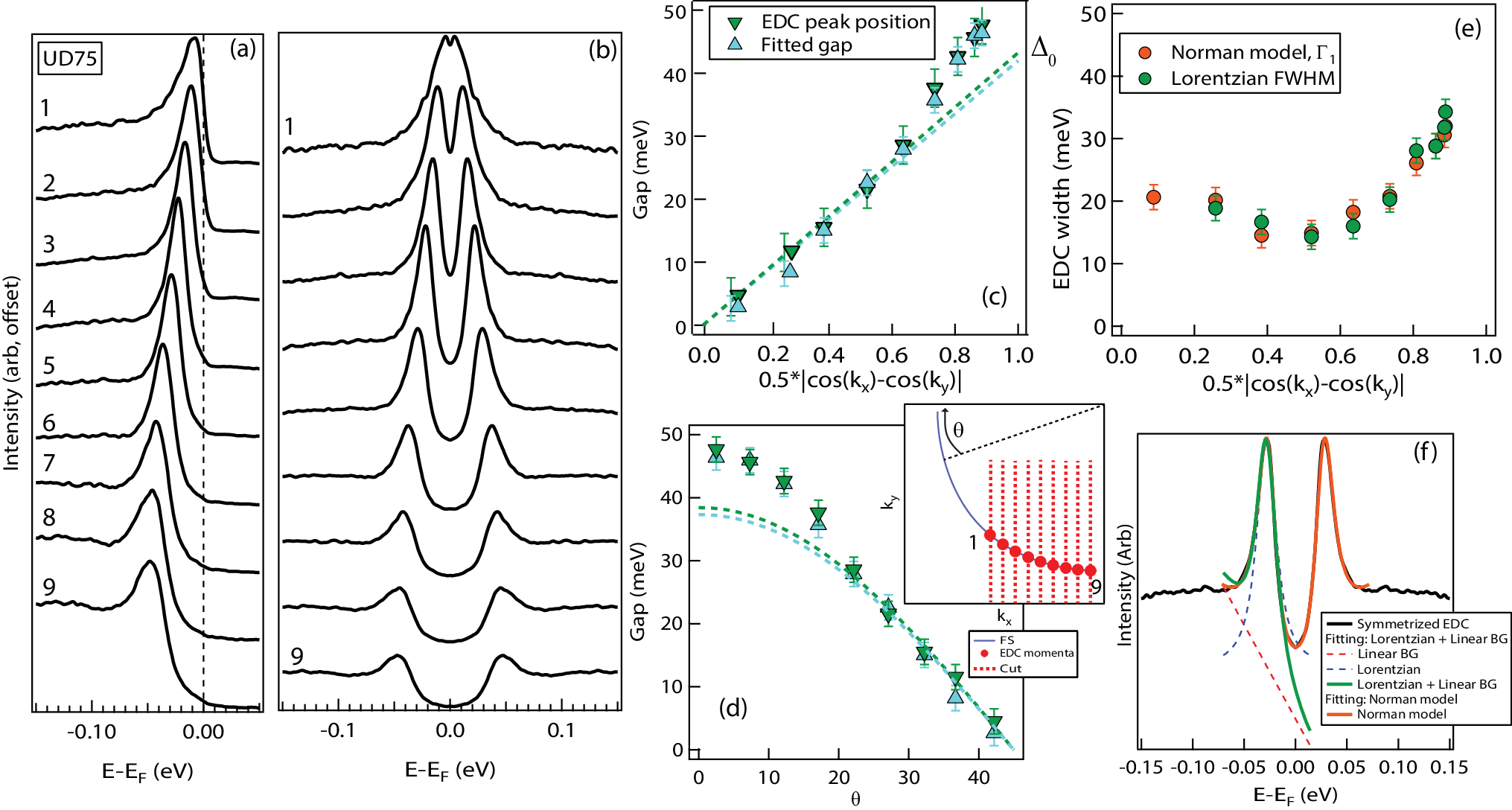}
\centering
\caption[QPs in UD75] {\label{Fig 1: UD75}  Antinodal EDCs and deviation from simple \textit{d}-wave form in sample whose doping is controlled only by oxygen annealing without any cation substitution. (a) Raw EDCs at k$_F$ for UD75 from near the node (1, top) to the antinode (bottom) (b) Symmetrized EDCs at k$_F$, same cuts as (a). (c) Gap as a function of the simple \textit{d}-wave form, represented by 0.5*$|\cos(k_x)-\cos(k_y)|$.  Gap defined from symmetrized EDCs in two ways: 1) fitting to assumed model (green) \cite{Symmetrization_Norman_model} and 2) from energy of EDC peak (cyan).  Dashed lines are linear fits to the first 5 data points.  (d) Gap as a function of Fermi surface angle, $\theta$, defined in inset.  Dashed lines are fits of first 5 data points to $\Delta(\theta)=\Delta_0\cos(2\theta)$ and color definitions are same as panel (c).  Inset: Fermi crossings for EDCs in (a)-(b) and gap fits in (c)-(d).  Dashed lines represent cut geometry of experiment. (e) EDC width as a function of simple \textit{d}-wave form, determined by two different methods: (1)the $\Gamma_1$ parameter in the Norman model \cite{Symmetrization_Norman_model} which defines the EDC width (orange) and (2) Lorentzian FWHM when a portion of the EDC is fit to a Lorentzian plus linear background (green). (e) Both EDC peak width fitting methods, shown for cut 5.}
\end{figure*}
\section{Data}
Fig. \ref{Fig 1: UD75} shows energy distribution curves (EDCs) at the Fermi momentum, k$_F$, for UD75, as well as gaps derived from those data.  Notably, the T$_c$ of this sample was achieved only by oxygen annealing, and no cation substitution.  Quasiparticles (QPs) are observed over the entire Fermi surface (FS), from the node to the antinode.  Symmetrization, given by $I(k_F,\omega)+I(k_F,-\omega)$, is used to remove the Fermi-Dirac cutoff, assuming particle-hole symmetry, which should be valid assumption in the superconducting state. The gap at each momentum is determined in two ways in Fig. \ref{Fig 1: UD75}(d): by fitting symmetrized EDCs to a minimal model \cite{Symmetrization_Norman_model} and from the peak positions of symmetrized EDCs.  Both yield results within error bars of one another.  The momentum dependence of the superconducting gap is illustrated by plotting as a function of the simple \textit{d}-wave form, expressed as 0.5*$|\cos(k_x)-\cos(k_y)|$.  This expression is equal to zero at the node (along the (0,0) to ($\pi$,$\pi$) line) and close to one at the antinode (where the FS meets the Brillouin zone boundary).  A superconducting gap is said to follow a simple \textit{d}-wave form if all data points fall on a straight line when plotted in terms of 0.5*$|\cos(k_x)-\cos(k_y)|$.  By this criterion, the gap function in Fig. \ref{Fig 1: UD75}(d) does not follow a simple \textit{d}-wave form over the entire FS.  In the near-nodal region, the gap function does follow a simple \textit{d}-wave form, and the linear fit to the first 5 data points is shown by dashed lines for both methods of determining gaps.  In the antinodal region, measured gaps are larger than the trend implied by the dotted lines, indicating a deviation of the gap function from a simple \textit{d}-wave form. Another expression for the simple \textit{d}-wave form is $\cos(2\theta)$, and gaps are plotted in terms of the FS angle, $\theta$ in Fig. \ref{Fig 1: UD75}(d).  Again, fitting the five data points closest to the node to a simple \textit{d}-wave form yields a trend from which the antinodal points deviate.  These data refute the claim in Ref. \onlinecite{Zhao:UniversalFeatures} that a deviation from a simple \textit{d}-wave form is only observed in cation substituted samples.  To emphasize the robustness of antinodal quasiparticles, the EDC width is plotted as a function of 0.5*$|\cos(k_x)-\cos(k_y)|$ in Fig. \ref{Fig 1: UD75}(e).  Two methods of quantifying the EDC width are shown.  The simplest method is to assume a Lorentzian peak on a background which varies linearly with binding energy.  The second method is from the so called single-particle scattering rate, $\Gamma_1$, from the same phenomenological model from which the gap was fit in panels (c)-(d).  The EDC width derived by both methods varies approximately by a factor of two around the Fermi surface, demonstrating that the EDC width does not diverge in the antinodal region.  This verifies that the criterion used to assess the gap is the same in the near-nodal and near-antinodal regions.  Notably, the EDC width in the antinodal region is smaller than the gap energy.  We note that the increase in EDC width near the node arises from momentum resolution which can broaden lineshapes in energy for a dispersive band.

\begin{figure}[!]
\includegraphics [type=eps,ext=.eps,read=.eps,clip, width=3 in]{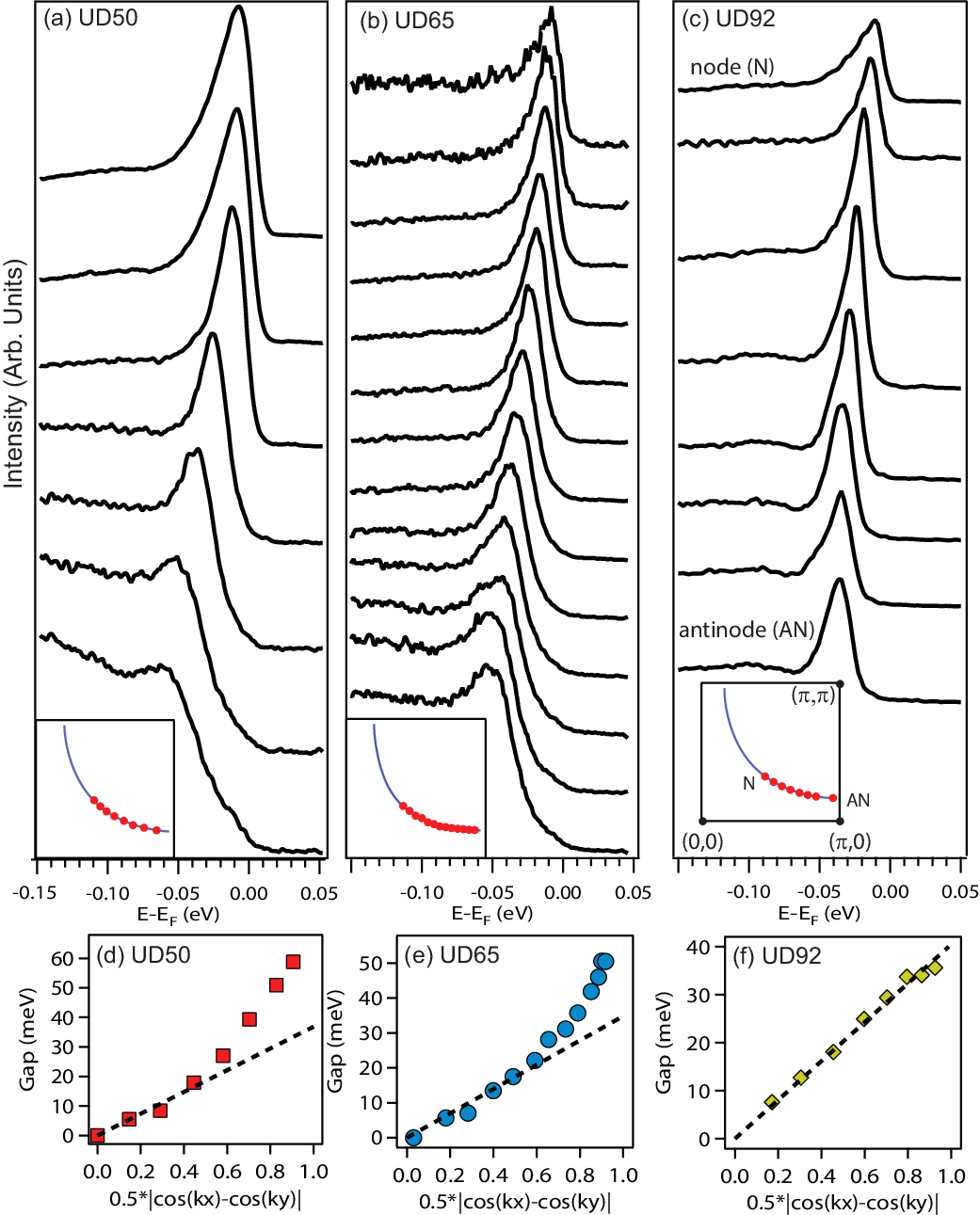}
\centering
\caption[Ubiquitous QPs] {\label{Fig 2: Ubiquitous QPs} Ubiquitous QPs all around the FS for various compositions and corresponding gap functions (a)-(c) EDCs at k$_F$ from the node(top) to the antinode (bottom) for UD50, UD65, and UD92.  Corresponding fitted gaps are plotted in (d)-(f).  Dashed lines in (d)-(f) represents simple \textit{d}-wave form.}
\end{figure}
Fig. \ref{Fig 2: Ubiquitous QPs} shows EDCs around the Fermi surface for several samples of varying compositions together with their fitted gaps.  In the most underdoped sample, UD50, doping is achieved by substituting Y on the Ca site.  The the second most underdoped sample, UD65, an excess of Bi achieves lower T$_c$.  UD92 represents the as-grown doping of Bi$_2$Sr$_2$CaCu$_2$O$_{8+\delta}$.  In Fig. \ref{Fig 2: Ubiquitous QPs}, the two most underdoped samples clearly show QPs all around the Fermi surface together with a deviation of the gap function from a simple \textit{d}-wave form.  This together with an identical result for UD75 in Fig. \ref{Fig 2: Ubiquitous QPs} demonstrates that the conclusion of a universal simple \textit{d}-wave form of the superconducting gap in Ref. \onlinecite{Zhao:UniversalFeatures} is simply not correct.  We note that in the inner-plane of a trilayer cuprate, Bi$_2$Sr$_2$Ca2Cu$_3$O$_{10+\delta}$ (Bi-2223), a deviation of the gap function from a simple \textit{d}-wave form is also accompanied by quasiparticles \cite{Ideta:EnhancedSC_Gaps}.  As an aside, recent high-resolution experiments showed subtle deviations from a simple \textit{d}-wave form in UD92 in the portion of the FS intermediate between the node and antinode, and this behavior persisted up to p$=$0.19 \cite{Vishik:PNAS, anzai:RelationNodalAntinodalGap}.

\begin{figure}[!]
\includegraphics [type=eps,ext=.eps,read=.eps,clip, width=2.75 in]{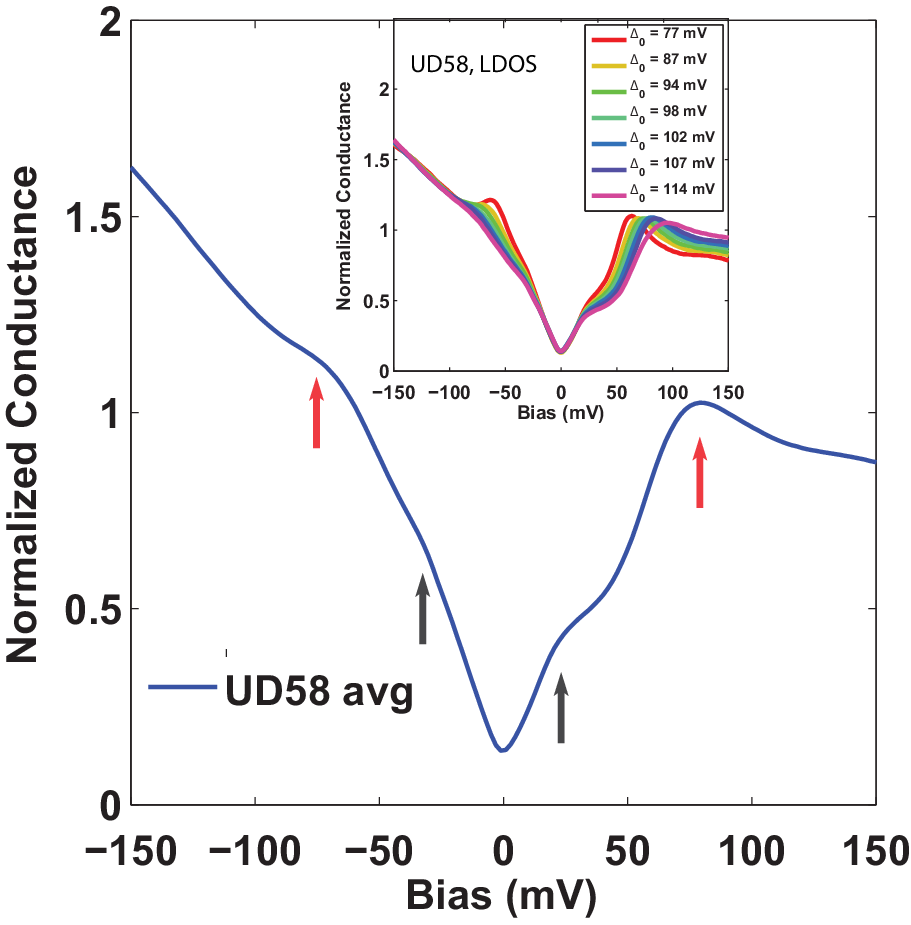}
\centering
\caption[STM] {\label{Fig 3: STM} STS spectrum, UD58 (Bi$_2$Sr$_2$(Ca,Dy)Cu$_2$O$_{8+\delta}$), averaged over 25nm $\times$ 25nm field of view.  Adapted from Ref. \onlinecite{Pushp:UniversalNode}.  Black and red arrows mark the energy scales of the smaller and larger gap. Measurement temperature is 20K.  Inset shows local density of states (LDOS) averaged over 5$\%$ of the scan area, with gap size of each local spectrum indicated. }
\end{figure}

The ARPES results shown in Figs. \ref{Fig 1: UD75} and \ref{Fig 2: Ubiquitous QPs} are in agreement with other spectroscopies, including Raman and scanning tunneling spectroscopy (STS) which show different energy scales in the near-nodal and near-antinodal regions of momentum space implying a deviation from a simple \textit{d}-wave form \cite{Boyer:ImagingTwoGapsBi2201,Hufner:2gaps,Pushp:UniversalNode}.  This is illustrated with STS data in Fig. \ref{Fig 3: STM}, showing spectra for a UD58 sample.  Spectra averaged over a large field of view clearly show two energy scales, marked with black and red arrows.  If the gap function followed a simple \textit{d}-wave form, STS would only show one energy scale, and this is indeed what is observed in overdoped samples \cite{Pushp:UniversalNode}.  In the \textit{d}-wave superconducting state, energies close to zero bias voltage correspond to momenta near the node, so the lower energy scale (black arrow) corresponds to the near-nodal gap and the higher energy scale (red arrow) corresponds to the antinodal gap.  At higher bias voltage, STS shows local inhomogeneity on several-nanometer length scales \cite{Kapitulnik:DOS_modulationSTM,Pan:MicroscopicInhomogeneity}, but crucially, the two distinct energy scales are clearly visible in local spectra as well as the spectrum averaged over a large field of view.  The averages spectrum is most relevant for comparisons to ARPES where the beam spot size is usually larger than 100 $\mu$m.  STS data suggest that a deviation from a simple \textit{d}-wave form is a generic feature of underdoped cuprates, rather than an anomaly observed by a single experimental technique.

\section{Discussion}

\begin{figure}[!]
\includegraphics [type=eps,ext=.eps,read=.eps,clip, width=2.5 in]{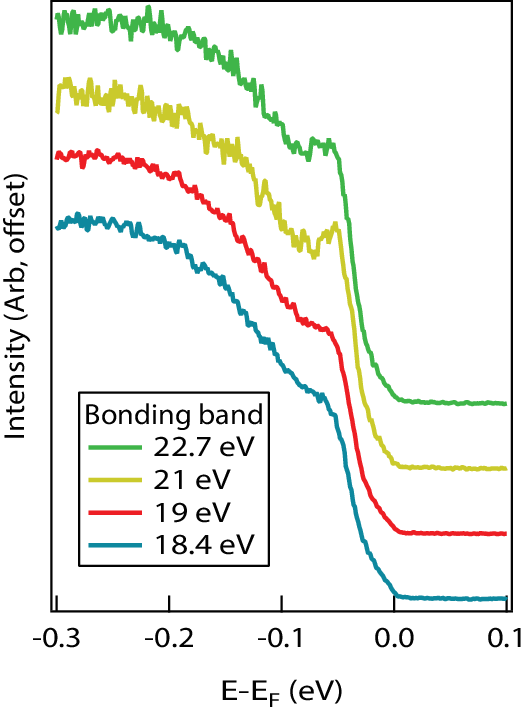}
\centering
\caption[matrix elements] {\label{Fig 3: matrix elements} EDCs at k$_F$ at the antinode (bonding band) for UD55.  Curves for different incident photon energies are offset vertically for clarity.  }
\end{figure}

What might be the cause of the differing conclusions from our data and from Ref. \onlinecite{Zhao:UniversalFeatures}?  A key point which must be clarified is that the absence of a quasiparticle is not as conclusive as the presence of a quasiparticle.  In particular, matrix element effects \cite{Campuzano:ARPES_Review,ARPES_Review,Bansil:MatrixElements2004} arising from a poor choice of experimental configuration (polarization, cut geometry, photon energy) can make a quasiparticle become less apparent in the spectrum.  To illustrate the effects of matrix elements, Fig. \ref{Fig 3: matrix elements} shows EDCs at the bonding band k$_F$ for UD55 taken with several different photon energies.  Data were taken on a single sample and a single cleave with beam polarization parallel to the Cu-O bond direction.  The quasiparticle peak is visible with 22.7 eV photons, enhanced with 21 eV photons, and greatly suppressed when measurements are done with 19 eV photons or 18.4 eV photons.  For the UD75 and UD92 samples, 22.7 eV photon energy enhances the intensity of the bonding band relative to the antibonding band, which is why this photon energy is commonly chosen for for experiments on Bi-2212 near optimal doping \cite{Lee:twoGapARPES_TDep}. However, for UD55, 22.7 eV does not yield the optimal cross section.  Similarly, 18.4 eV photon energy is a common choice to enhance the antibonding band in overdoped Pb-doped samples, but for UD55, this photon energy yields poor cross section for both the bonding and antibonding bands. In deeply underdoped Y-Bi-2212, the optimal cross-section was found at 19eV photon energy measured in the second Brillouin zone \cite{Tanaka:twoGapARPES_dopingDep}.  Different experimental configurations must be thoroughly explored before concluding that a quasiparticle is absent. 

Another difference between experiments is that the ones highlighted in this manuscript were typically performed with energy resolutions of better than 10 meV, while those of Zhao \textit{et al} were performed with energy resolutions of 15-20 meV.  A poorer resolution can affect the EDC peak position, particularly in the near-nodal region where bands are more dispersive and gaps are smaller.  This can diminish the visibility of details of the momentum dependence of the gap, such as a deviation from a simple \textit{d}-wave form, in the superconducting state which are clearly visible in experiments performed with better resolution.  Simulations of energy and momentum resolution effects are shown in Supplementary Information.

The work in Ref. \onlinecite{Zhao:UniversalFeatures} brought up an important question about the effects of chemistry on the momentum-resolved electronic structure of cuprates, and it is crucial that future ARPES experiments address this with adequate rigor.  This is an important avenues for understanding the mechanism of high temperature superconductivity: in single-layer cuprates. T$_{c,max}$ varies by more than a factor of two depending on the composition of the charge reservoir layers between the CuO$_2$ planes and the types of disorder which are present \cite{Eisaki:ChemicalInhomogeneityAndTypesOfDisorder}.  ARPES has addressed this issue in single-layer Bi$_2$Sr$_{1.6}$L$_{0.4}$CuO$_{6+\delta}$ (L $=$ La, Nd, Eu, Gd), and Gd doping has been shown to lower T$_c$ substantially \cite{Fujita:Disorder2201}, accompanied by a loss of antinodal coherence and an enhanced antinodal gap \cite{Hashimoto:OutOfPlaneDisorder}.  Eu doping also suppresses T$_c$, and has been shown to increase T* \cite{Okada:JPSJ_disorder_T_star}.  In Bi-2212, crystal-growth studies have shown that Sr-site disorder, particularly, excess Bi on the Sr site, suppresses T$_c$, although not as acutely as in Bi-2201 \cite{Hobou:EnhancementTcBi2212}.  Bi-2212 samples with a stochiometric composition have a maximum T$_c$ of 94K, but typical samples have an excess of Bi, which brings the maximum T$_c$ down to 89-92K \cite{Eisaki:ChemicalInhomogeneityAndTypesOfDisorder}.  By doping 8$\%$ Y on the Ca site, it is possible to raise the maximum T$_c$ of Bi-2212 to 96K, by stabilizing growth of samples with stochiometric Bi and Sr concentrations.  The fact that cation substitution is \textit{required} to maximize the T$_c$ of Bi-2212 refutes claims that cation substituted samples are more disordered.  However, the effect this has on ARPES observables--gaps, dispersions, and spectral weight--has not been explored systematically.  This is crucial for interpreting ARPES data going forward.  However, this should not be confused by the claims of Ref. \onlinecite{Zhao:UniversalFeatures} that attribute the shape of the gap function to chemistry effects.  Data collected from similar samples as in Ref. \onlinecite{Zhao:UniversalFeatures} with better resolution and optimized matrix element detected clear QP peak, questioning the claim that the QP are suppressed in deeply underdoped samples due to this chemistry effect.  Additionally, similar results are observed in other cuprates by a number of research groups \cite{Kaminski:2gapBi2201,Ideta:EnhancedSC_Gaps,Nakayama:DopingDepYBCO}.  The work presented in this manuscript reaffirms the intrinsic nature of a gap function that deviates from simple \textit{d}-wave form in underdoped cuprates, suggesting a pseudogap origin of this observation.

\begin{acknowledgements}
We acknowledge helpful discussions with Ali Yazdani and Colin Parker.  This work is supported by DOE Office of Basic Energy Science, Division of Materials Sciences.
\end{acknowledgements}


\end{document}


\title{Ubiquitous antinodal quasiparticles and deviation from simple \textit{d}-wave form in underdoped Bi-2212: Supplementary materials}
\maketitle

\renewcommand{\thesection}{S.\arabic{section}}
\renewcommand{\thesubsection}{\thesection.\arabic{subsection}}

\renewcommand{\bibnumfmt}[1]{[S#1]}
\renewcommand{\citenumfont}[1]{S#1}

\begin{figure*}
\includegraphics [type=eps,ext=.eps,read=.eps,clip, width=6.0 in]{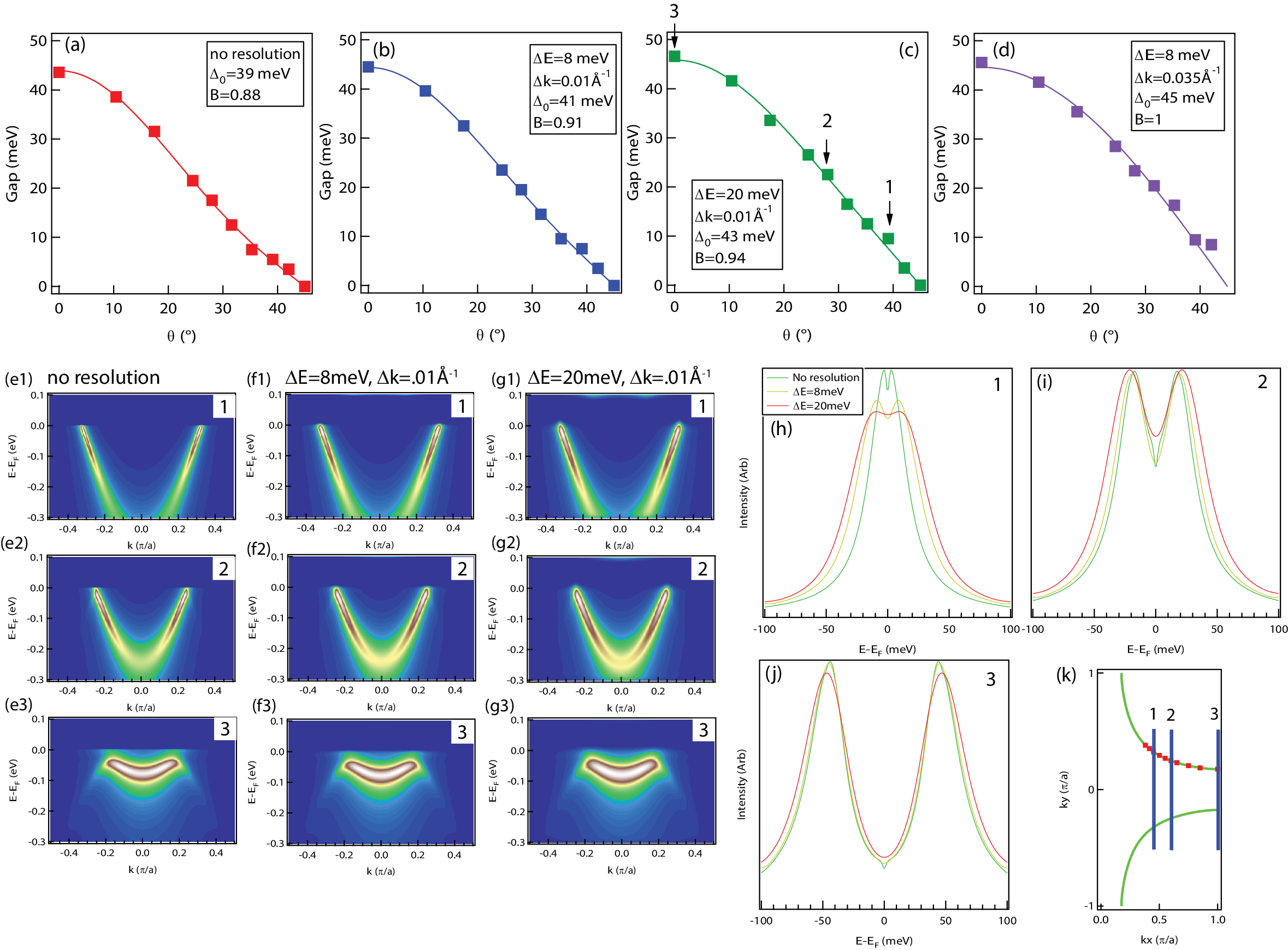}
\centering
\caption[resolution summary] {\label{Fig 1: resolution summary} (a)-(d) Gap vs Fermi surface angle for various choices of energy and momentum resolution in simulation.  Fits are to a \textit{d}-wave gap with a higher harmonic term: $\Delta(\theta)=\Delta_0*[\cos(2\theta)+(1-B)*\cos(6\theta)]$.  Panel (c) marks the three cuts which are detailed in subsequent panels. (e)-(g) simulated data with cuts taken parallel to the Brilliouin zone boundary. In (f)-(g), data has been convolved with a resolution function. (h)-(j) EDCs at k$_F$ with varying resolution. (k) Fermi surface schematic, Fermi crossings used in (a)-(d), and cuts used in (e)-(g).}
\end{figure*}

To assess the effects of instrument resolution on the measured gap function, we simulate spectra using a model spectral function proposed by Kordyuk et al. \cite{KordyukModel}:
\begin{equation}
\label{ImS ReS Kordyuk}
\Sigma"(\mathbf{k},\omega)=\sqrt{(\alpha\omega)^2+(\beta T)^2}
\end{equation}
The simulations in Fig. S\ref{Fig 1: resolution summary} use the values $(\alpha, \beta, T)=(0.2, 9.3, 10K)$ which produces spectra comparable to experiments. An additional energy-independent momentum broadening of 0.006 $\AA^{-1}$ is included for better agreement with  data.  The band structure is taken to be a tight binding model with parameters
 \begin{equation}
 (\mu, t, t', t'', t''', t'''')=(0.091, -0.72, 0.063,-0.17, -0.015, 0.073)
 \end{equation}
The resolution function is taken to be a two-dimensional Gaussian in energy and momentum with values chosen to be relevant to experiments.  Momentum resolution perpendicular to the analyzer slit is neglected.

Instrumental resolution not only broadens spectra, but it can also move the peak position of energy distribution curves (EDCs) slightly. Because there is no spectral weight above E$_F$ in a low-temperature experiment, applying a convolution operation to a band which crosses E$_F$ will tend to push weight asymmetrically near E$_F$.  In the cuprates, the near-nodal region is most strongly affected by resolution effects because the Fermi velocity is larger.  This is seen in Fig. S\ref{Fig 1: resolution summary}(h)-(j) which shows symmetrized EDCs at k$_F$ for three cuts indicated in panel (k).  Closer to the node, instrumental energy resolution pushes the EDC peak to higher binding energy, but this effect is less pronounced at the antinode where bands are less dispersive.

The simulations are summarized in Fig. S\ref{Fig 1: resolution summary}(a)-(d) which shows gaps around the Fermi surface for different choices of instrument resolution.  The gap function is deliberately chosen to deviate from a simple \textit{d}-wave form, and is instead described by a form with a higher harmonic: $\Delta(\theta)=\Delta_0*[\cos(2\theta)+(1-B)*\cos(6\theta)]$.  The parameter B quantifies the degree of deviation from a simple \textit{d}-wave form, with B$=$1 signifying a simple \textit{d}-wave form and smaller values indicating greater deviation from a simple \textit{d}-wave form.  With increasing energy resolution and identical momentum resolution, the B parameter approaches 1, indicating decreasing deviation from a simple \textit{d}-wave form.  This may be one of the reasons Ref. \onlinecite{Zhao:UniversalFeatures} reports a simple \textit{d}-wave form for all dopings.  When the deviation from a simple \textit{d}-wave form is subtle, poorer resolution can obscure this effect.  Increasing the momentum resolution (Fig. S\ref{Fig 1: resolution summary}(d)) can also smooth out subtle features in the gap function.

\makeatletter
\makeatletter \renewcommand{\fnum@figure}
{\figurename~S\thefigure}
\makeatother

\clearpage
\bibliography{supplement_refs}